\documentstyle[epsfig]{aipproc}
\newcommand{\emet}{{\em et al.}}

\begin{document}

\title{{\small Invited Paper presented at 3rd International Symposium on 
\hfill TRI--PP--00--18 \\ 
\vspace*{-0.15cm}
Symmetries in Subatomic Physics, Adelaide, March 13--17 \hfill April 2000\\}
\vspace*{0.2cm}
Constraints on a Parity-even/Time-Reversal-odd Interaction
\footnote{Work supported in part by the Natural Sciences and Engineering 
Research Council of Canada.}}
\author{Willem T.H. van Oers}
\address{
Department of Physics, University of Manitoba, Winnipeg, MB, 
Canada R3T 2N2\\
and\\
TRIUMF, 4004 Wesbrook Mall, Vancouver, BC  Canada V6T 2A3}
\maketitle

\begin{abstract}
    Time-Reversal-Invariance non-conservation has for the first time been
unequivocally demonstrated in a direct measurement, one of the results of
the CPLEAR experiment. What is the situation then with regard to
time-reversal-invariance non-conservation in systems other than the neutral
kaon system? Two classes of tests of time-reversal-invariance need to be
distinguished: the first one deals with parity violating
(P-odd)/time-reversal-invariance non-conserving (T-odd) interactions, 
while the second one deals with P-even/T-odd interactions 
(assuming CPT conservation this implies C-conjugation non-conservation). 
Limits on a
P-odd/T-odd interaction follow from measurements of the electric dipole
moment of the neutron. This in turn provides a limit on a P-odd/T-odd
pion-nucleon coupling constant which is $10^{-4}$ times the weak
interaction strength. Limits on a P-even/T-odd interaction are much less
stringent. The better constraint stems also from the measurement of the
electric dipole moment of the neutron. Of all the other tests, measurements
of charge-symmetry breaking in neutron-proton elastic scattering provide
the next better constraint. The latter experiments were performed at TRIUMF (at
477 and 347 MeV) and at IUCF (at 183 MeV). Weak decay experiments (the
transverse polarization of the muon in $K^{+} \to \pi^{0} \mu^{+}
\nu_{\mu}$ and the transverse polarization of the positrons in polarized
muon decay) have the potential to provide comparable or possibly better
constraints.
\end{abstract}


\section*{Introduction}

    Time-reversal-invariance non-conservation has for the first time been
unequivocally demonstrated in a direct measurement in the CPLEAR 
experiment. [1]
The experiment measured the difference in the transition probabilities
$P(\overline{K}^{0} \to K^{0})$ and $P(K^{0} \to \overline{K}^{0})$. 
Assuming CPT conservation
but allowing for a possible breaking of the $\Delta$S = $\Delta$Q rule, the
result obtained for $A_T$
\begin{eqnarray}
A_T = \frac{R(\overline{K}^{0} \to K^{0}) - R(K^{0} \to
    \overline{K}^{0})}
{R(\overline{K}^{0}  \to K^{0}) + R(K^{0} \to \overline{K}^{0})} 
= [6.6 \pm 1.3{\rm (stat.)} \pm 1.0{\rm (syst.)}] \times 10^{-3}
\end{eqnarray}
is in good agreement with the measure of CP violation in neutral kaon
decay.  A more recent reported result is a large asymmetry in the
distribution of $K_L \to \pi^+ \pi^- e^+ e^-$ events in the CP-odd/T-odd
angle $\phi$ between the decay planes of the $\pi^+ \pi^-$ and $e^+ e^-$
pairs in the $K_L$ centre of mass system. The overall asymmetry found was
[13.6~$\pm$~2.5(stat.)$\pm$ 1.2(syst.)]\%. [2] The question then is: what is
the situation with regard to time-reversal-invariance in systems other than
the kaon system?

    Tests of time-reversal-invariance can be distinguished as belonging to two
classes: the first one deals with P-odd/T-odd interactions, while the second
one deals with P-even/T-odd interactions (assuming CPT conservation this
implies C-conjugation non-conservation). But it should be noted that
constraints on these two classes of interactions are not independent since the
effects due to P-odd/T-odd interactions may also be produced by 
P-even/T-odd
interactions in conjunction with Standard Model parity violating radiative
corrections. The latter can occur at the $10^{-7}$ level and consequently could
present a limit on the constraint of a P-even/T-odd interaction, 
derived from experiment. Limits on a P-odd/T-odd interaction follow from
measurements of the electric dipole moment of the neutron (which currently
stands at $<6 \times 10^{-26}$ e.cm [95\% C.L.]). This provides a limit on a
P-odd/T-odd pion-nucleon coupling constant which is less than $10^{-4}$ times
the weak interaction strength. Measurements of $^{129}$Xe and $^{199}$Hg
($<8 \times 10^{-28}$ e.cm [95\% C.L.]) give similar constraints. [see Ref.~3]

    Experimental limits on a P-even/T-odd interaction are much less stringent.
Following the standard approach of describing the nucleon-nucleon 
interaction in
terms of meson exchanges, it can be shown that only charged rho-meson exchange
and A$_1$-meson exchange can lead to a P-even/T-odd interaction. [4] The better
constraints stem from measurements of the electric dipole moment of the neutron
and next from measurements of charge symmetry breaking in neutron-proton
($n$-$p$) elastic scattering. All other experiments, like gamma decay experiments
[5], detailed balance experiments [6], polarization - analyzing power 
difference
measurements [7], and five-fold correlation experiments with polarized incident
nucleons and aligned nuclear targets [8], have been shown to be at least an
order of magnitude less sensitive. Haxton, Hoering, and Musolf [3] have deduced
constraints on a P-even/T-odd interaction from nucleon, nuclear, and atomic
electric dipole moments with the better constraint coming from the electric
dipole moment of the neutron. In terms of a ratio to the strong rho-meson
nucleon coupling constant, they deduced for the P-even/T-odd rho-meson
nucleon coupling: $|\overline{g_\rho}| < 0.53 \times 10^{-3} \times 
|f^{\rm DDH}_\pi/f^{\rm meas.}_\pi|$.
But the ratio of the theoretical to the experimental value of $f_\pi$ may be
as large as 15! [9] However, constraints derived from one-loop contributions to
the electric dipole moment of the neutron exceed the two-loop limits by more
than an order of magnitude and are much more stringent. [10] However, a
translation in terms of coupling strengths in the hadronic sector still
needs to be made.

    It is very difficult to accommodate a P-even/T-odd interaction in the
Standard Model. It requires C-conjugation non-conservation, which cannot be
introduced at the first generation quark level. It can neither be introduced
into the gluon self-interaction. Consequently, one needs to consider
C-conjugation non-conservation between quarks of different generations and/or
between interacting fields. [11]

\section*{Charge Symmetry Breaking in Neutron Proton Elastic Scattering}

    Charge symmetry breaking (CSB) in neutron-proton elastic scattering
    manifests itself as a non-zero difference of the neutron $(A_n)$ and
    proton $(A_p)$ analyzing powers, $\Delta{A} = A_n - A_p = 2 \times
    [{\rm Re}(b^{*}f) +
{\rm Im}(c^{*}h)]/\sigma_0$ .  Here the complex amplitude $f$ is charge
symmetry breaking, while the complex amplitude $h$ is both charge symmetry
breaking and time-reversal-invariance non-conserving. The complex
amplitudes $b$ and $c$ belong to the usual five $n$-$p$ scattering
amplitudes and $\sigma_0$ is the unpolarized differential cross
section. The three precision experiments performed (at TRIUMF at 477 MeV
[12] and at 347 MeV [13], and at IUCF at 183 MeV [14]) have unambiguously
shown that charge symmetry is broken and that the results for $\Delta{A}$
at the zero-crossing angle of the average analyzing power are very well
reproduced by meson exchange model calculations (see Fig.~1). A
P-even/T-odd interaction produces a term in the scattering amplitude which
is simultaneously charge symmetry breaking (the complex amplitude $h$ in
the expression above). Thus, Simonius [15] deduced an upper limit on a
P-even/T-odd CSB interaction from a comparison of the experimental results
with the theoretical predictions for the three $n$-$p$ CSB experiments. The
upper limit so derived is $|\overline{g}_\rho| < 6.7 \times 10^{-3}$ [95\%
{\rm C.L.}]. This is therefore comparable to the upper limit deduced from
the electric dipole moment of the neutron, taking present experimental
limits of $f_\pi$, and is considerably lower than the limits inferred from
direct tests of a P-even/T-odd interaction. For instance the detailed
balance experiments give a limit of $|\overline{g}_\rho| < 2.5 \times
10^{-1}$ [see Ref.~8], while measurements of the five-fold correlation
parameter $A_{y,xz}$ in polarized neutron transmission through nuclear
spin-aligned $^{165}$Ho give a limit of $|\overline{g}_\rho| < 5.9 \times
10^{-2}$ even though the measured value of $A_{y,xz}$ was $(8.6 \pm 7.7)
\times 10^{-6}$. [8] It is effectively only the valence proton in
$^{165}$Ho which contributes to $A_{y,xz}$. Even though it is inconceivable
in the Standard Model to account for a P-even/T-odd interaction, there is a
need to clarify the experimental situation by providing a better
experimental result.

\begin{figure}[t!]
\centerline{\epsfig{figure=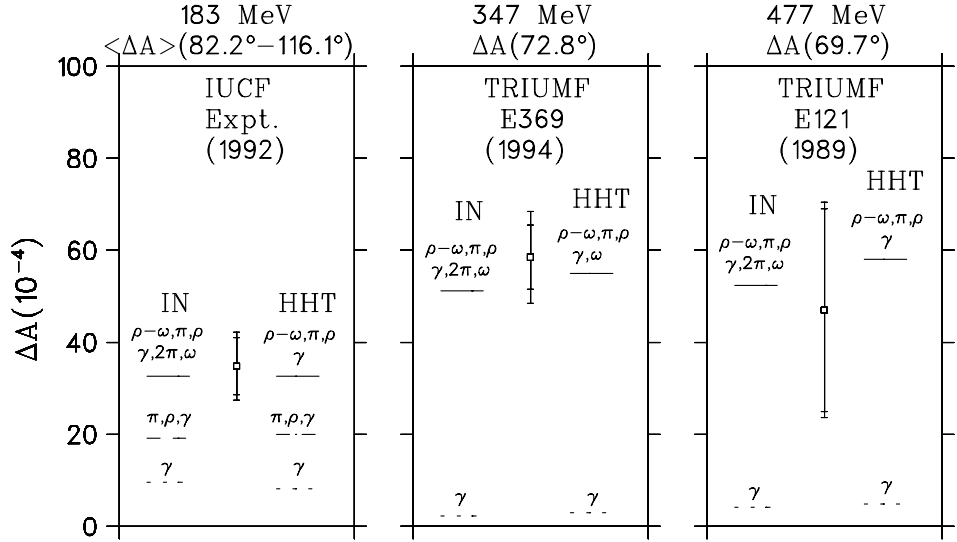,height=0.5\linewidth}}
\caption{Experimental results of $\Delta{A}$ at the zero-crossing angle at
incident neutron energies of 183, 347, and 477 MeV compared with
theoretical predictions of Iqbal and Niskanen, and Holzenkamp, Holinde, and
Thomas. The inner error bars present the statistical uncertainties; the
outer error bars have the systematic uncertainties included (added in
quadrature). For further details see Ref.~13.}
\vspace*{-2mm}
\label{fig1}
\end{figure}

    Such an experimental constraint may be provided by an improved upper
limit on the electric dipole moment of the neutron. In fact a new
measurement with a sensitivity of $4 \times 10^{-28}$ e.cm has been proposed at
the Los Alamos Neutron Science Center. [16] Performing an improved $n$-$p$
elastic scattering CSB experiment also appears to be a very attractive
possibility. One can calculate with a great deal of confidence the
contributions to CSB due to one-photon exchange and due to the $n$-$p$ mass
difference affecting charge one-pion and rho-meson exchange. Furthermore,
one can select an energy where the $\rho^\circ-\omega$ meson mixing contribution
changes sign at the same angle where the average of the analyzing powers
$A_n$ and $A_p$ changes sign and therefore does not contribute to
$\Delta{A}$.
This occurs at a neutron energy of 320 MeV and is caused by the
particular interplay of the $n$-$p$ phase shifts and the form of the
spin/isospin operator connected with the $\rho^\circ-\omega$ mixing term. But
also the one-photon exchange term changes sign at about the same angle
at 320 MeV. The contribution due to two-pion exchange with an
intermediate $\Delta$ is expected to be less than one tenth of the
overall CSB effect, essentially determining an upper limit on the
theoretical uncertainty (see Fig.~2). [17] It has been shown that
simultaneous $\gamma$-$\pi$ exchanges can only contribute to $\Delta{A}$
through second order processes and can therefore be neglected. [18]
Also the effects of inelasticity are negligibly small at 320 MeV.
It appears therefore well within reach to reduce the theoretical
uncertainty in the comparison of experiment and theory. Subtracting the
calculated $\Delta{A}$ from the measured $\Delta{A}$ permits establishing an
upper limit on a P-even/T-odd/CSB interaction.

\begin{figure}[t!]
\centerline{\epsfig{figure=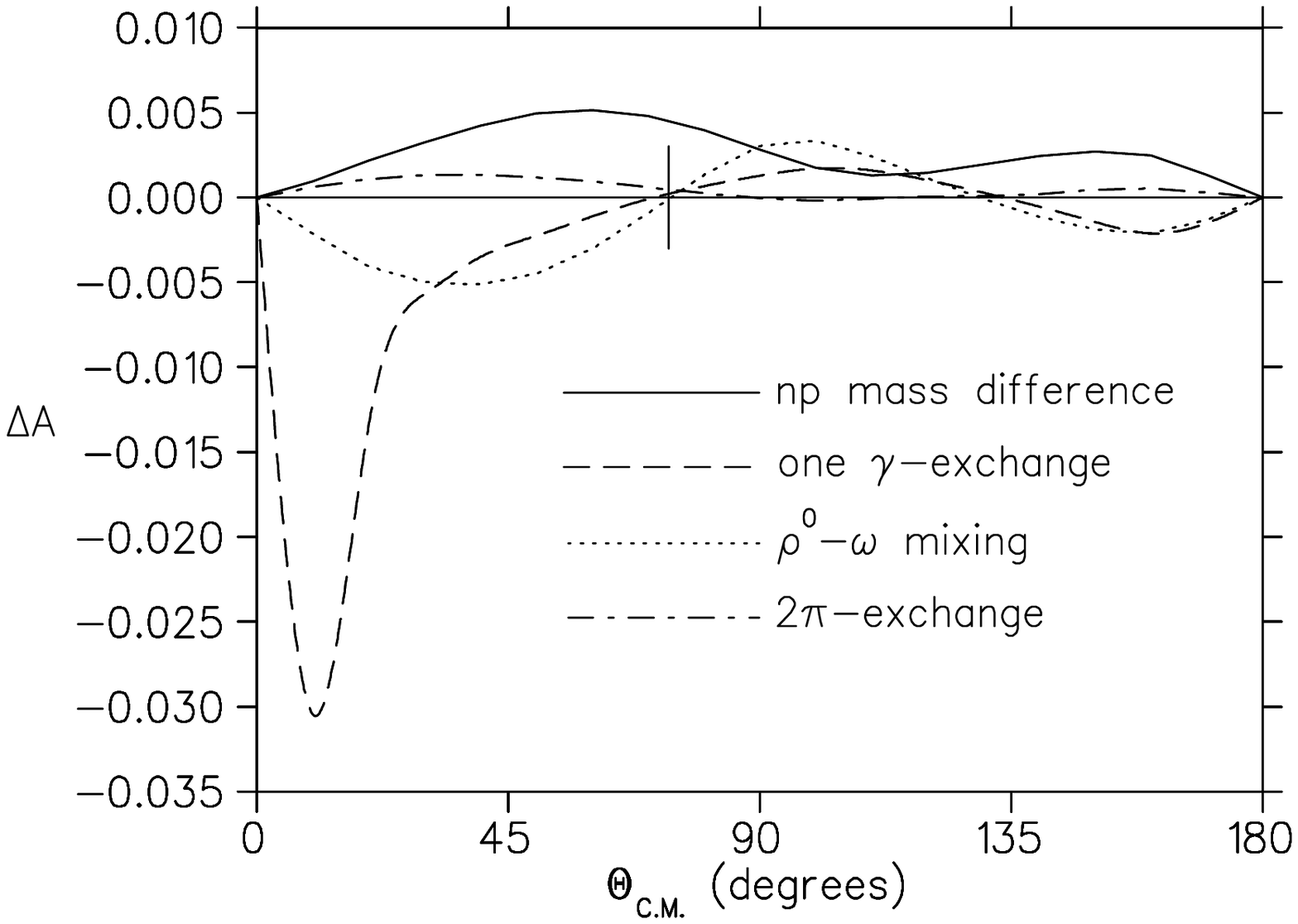,height=0.4\linewidth}}
\centerline{\epsfig{figure=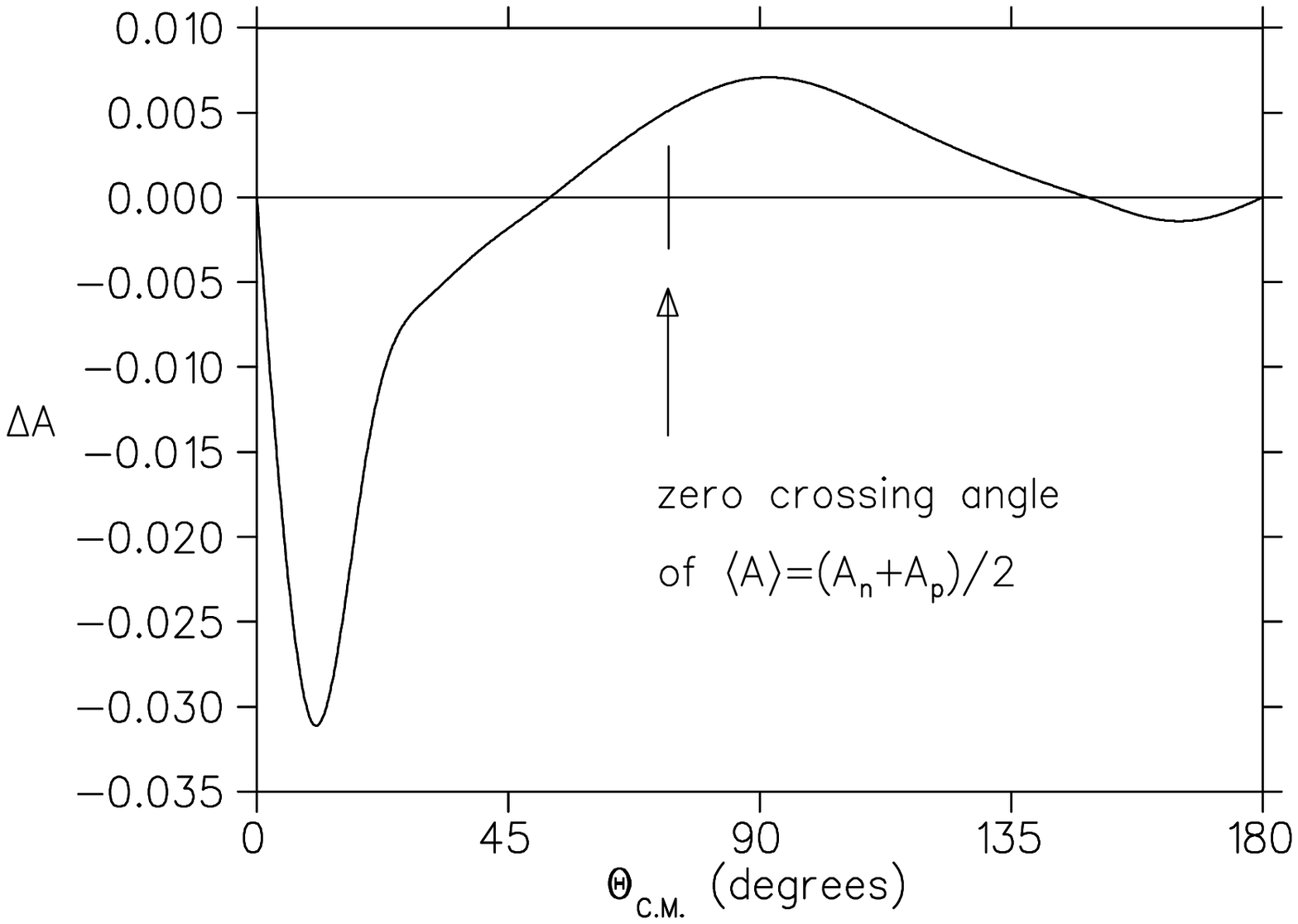,height=0.4\linewidth}}
\caption{Angular distributions of the different contributions to $\Delta{A}$
at an incident neutron energy of 320 MeV. (Ref.~17) Note that the $\rho^0$ -
$\omega$ mixing contribution passes through zero at the same angle as the
average of $A_n$ and $A_p$ (vertical bar). The lower part of the figure gives
the total $\Delta{A}$ angular distribution.}
\label{fig1}
\vspace*{-2mm}
\end{figure}

    In the TRIUMF CSB experiments polarized neutrons were scattered from
unpolarized protons and vice versa. The polarized (or unpolarized)
neutron beam was obtained using the ($p,n$) reaction with a 369 (and 497)
MeV polarized (or unpolarized) proton beam incident on a 0.20 m long
LD$_2$ target. At these energies one makes use of the large
sideways-to-sideways polarization transfer coefficient $r_t$ at 9$^\circ$
in the lab. The only difference in obtaining the polarized and
unpolarized 347 MeV neutron beams was turning off the pumping laser light
in the optically pumped polarized ion source (OPPIS). The polarized proton
target was of the frozen spin type with butanol beads as target material.
The same target after depolarization was used as the unpolarized proton
target. Great care was taken that the two interleaved phases of the
experiments were performed with identical beam and target parameters
except for the polarization states. At 347 MeV scattered neutrons and
recoiling protons were detected in coincidence in the c.m. angular range
53.4 to 86.9 degrees in two left-right symmetric detector systems. Rather
than measuring $A_n$ and $A_p$ directly (which would limit the accuracy
attainable by not having polarization calibration standards of the
required precision), the zero-crossings of $A_n$ and $A_p$ were determined
by fitting the partial angular distributions with polynomials, deduced/.
from $n-p$ phase shift analyses. The difference $\Delta{A}$ followed then by
multiplying the difference in the zero-crossing angles by the average
slope of the analyzing powers (the experiment measured the slope of $A_p$
at the zero-crossing angle, which is a good approximation for the
average slope at the zero-crossing angle and introduces a negligible
error). The execution of the experiments depended on a great deal of
simultaneous monitoring and online control measurements. Both the
statistical and systematic errors, obtained in the 347 MeV experiment,
can be considerably improved upon (by a factor three to four). With the
OPPIS developments which have taken place in the intervening years and
with a biased Na-ionizer cell it will be possible to obtain up to
50 $\mu$A of 342~MeV 80\% polarized proton beam incident on the neutron
production target (a factor of 50 increase in neutron beam intensity
at 320 MeV over the previous 347 MeV CSB experiment). In addition
various systematic error reducing improvements can be introduced in the
experimental arrangements and procedures. Such an experiment would
constitute a measurement of CSB in $n$-$p$ elastic scattering of
unprecedented precision of great value on its own and would
simultaneously provide a greatly improved upper limit on a P-even/T-odd
interaction.

\section*{Particle Decays}

    Searches for P-even/T-odd interactions are also made in particle decays,
e.g., in the decay $\mu^+ \to e^+ \nu_e \overline{\nu}_{\mu}$ and in the
decay $K^+ \to \mu^+ \pi^0 \nu_{\mu}$. A non-zero value of the muon polarization transverse to the decay plane
would be an indication of time-reversal-invariance non-conservation. An
experiment to measure the first decay is presently being executed at PSI. [19]
Several experiments have been performed to measure the transverse muon
polarization in both neutral and charged kaon decay. There is a unique
feature to the transverse muon polarization in that it does not have
contributions from the Standard Model at tree level and that higher order
effects are of order $10^{-6}$. When only one charged particle is present
in the final state, a final state interaction, which can mimic a
time-reversal-invariance breaking effect, is greatly reduced and is
estimated to occur only at the same level of $10^{-6}$. The more recent effort
of measuring the time-reversal-invariance non-conserving transverse muon
polarization is being done at KEK using a stopped $K^+$ beam. The
experiment reported a result for P$_T$~=~-0.0042 $\pm$ 0.0049(stat.) $\pm$
0.0009(syst.), based on the data taken in 1996 and 1997, which translates
into a value of Im{$\xi$} = -0.013 $\pm$ 0.016(stat.) $\pm$
0.003(syst.). [20] The quantity $\xi$ is defined as the ratio of two form
factors, $f_{+}(q^2)$ and $f_{-}(q^2)$, in the K$_{\mu 3}$ decay.   
Im$\xi$ must be equal to zero for time-reversal-invariance to hold. With
the data already in hand and with the approved data taking time, it is
anticipated to arrive at a statistical error of $\pm$0.008 in Im$\xi$. The
best previous experimental limits were obtained with both neutral and
charged kaons at the BNL-AGS. [21] A combination of both experimental
results provided a limit on the imaginary part of the hadron form factor,
Im$\xi$ = -0.010 $\pm$ 0.019. A new search for the time-reversal-invariance
non-conserving transverse muon polarization with in-flight decays of $K^+ \to
\mu^+ \pi^0 \nu_{\mu}$ was proposed at the BNL-AGS.
  [22] It was intended to obtain a sensitivity to the transverse muon
polarization of $\pm$0.00013, corresponding to a sensitivity to Im$\xi$ of
$\pm$0.0007. Similar searches for the time-reversal-invariance non-conserving
transverse $\tau$ polarization in B semileptonic decays, $B \to M\tau\nu_{\tau}$
are under consideration. Significant transverse $\tau$ lepton polarizations have
been predicted. Clearly, a non-zero value of the transverse muon polarization in
$K_{\mu 3}$ decay, or of the $\tau$ lepton in semi-leptonic $B$ decays would
constitute evidence for new physics.

\section*{Summary}

    The searches made so far for a P-even/T-odd interaction have 
resulted in only very modest constraints on such an interaction. Most
promising are the continued efforts to measure the electric dipole moment
of the neutron and secondly charge symmetry breaking in neutron-proton
elastic scattering at around 320 MeV.  But also measurements of transverse
lepton polarizations in $\mu$, $K$, and $B$ decays have the potential to
set better experimental limits on a P-even/T-odd interaction.


\end{document}